\documentstyle [12pt]{article}
\normalbaselines
\begin{document}

\begin{center}
{\large \bf  WHAT IS THE FORCE BETWEEN ELECTRONS?}

\vspace{3cm}

J. Sucher\\ [5mm]
{\it Center for Theoretical Physics and Department of Physics
University of Maryland, College Park, MD 20742, USA}\\[5mm]

\begin{abstract}

The concept of the potential acting between charged particles is
reviewed within the framework of quantum electrodynamics.  The
one-photon and two-photon exchange potentials $V_1 \gamma$ and $V_2
\gamma$ are discussed, with attention to the following features:  the
dependence of $V_2 \gamma$ on the choice of $V_1 \gamma$, the
distinction between FGI (Feynman gauge inspired) and CGI (Coulomb gauge
inspired) potentials, and the dependence of the large-r behavior on
these choices.  A summary of available results for two spin-0 particles,
for a spin-0 and a spin 1/2 particle and for two spin-1/2 particles is
given.  The application of such "scattering potentials" to the
calculation of energy levels of bound states is discussed.
\end{abstract}

\end{center}

Contents

I. Introduction and historical review

II. Scattering potentials: One-photon exchange
 
	A. Two spin-1/2 particles

	B. Two spin-0 particles

	C. A spin-1/2 and a spin-0 particle

	D. Remarks on the orbit-orbit interaction

III. Scattering potentials: Two-photon exchange

	A. Two spin-0 particles

	B. A spin-1/2 and a spin-0 particle

	C. Two spin-1/2 particles

IV. Scattering potentials and bound states

\section{Introduction and historical review}

It is a great pleasure for me to be able to contribute to this volume
in honor of my good friend, Ingvar Lindgren.
A large part of Ingvar's illustrious career has been devoted to work
in theoretical atomic physics.  Since the structure and properties of atoms
are largely determined by the nature of the interaction between its electrons,
I thought this would be a nice occasion to present a discussion of this topic.
I will focus on aspects which are of especial importance for the relativistic
theory of atomic structure, to which Ingvar has made so many contributions.
Because he has had an amazingly large number of graduate students
(over 60!), there is a good chance that some of them will find something here
which Ingvar would think they ought to know. (Note: The exclamation point after
60 indicates astonishment, not ``factorial'' - even Ingvar has his limitations.)

The question posed in the title may be short, but it does not
have a snappy answer.  If one asks ``What is the force
between planets?'', everybody knows that it is given with great accuracy by
Newton's law of universal gravitation.  But for electrons, the story is much
more complicated and indeed the question itself must be recast.  There are at
least three reasons for this: (i) physical systems involving electrons must be
described by quantum mechanics (QM), (ii) in many such systems, the relative
velocities are either not very small compared to \it c \rm and/or high accuracy is of
interest, so special relativity must be used, and (iii) electrons have spin.

With regard to (i), of course in nonrelativistic
classical mechanics the concept of force and,
in particular, the concept of the force between particles is well-defined;
if the force is conservative, so is the concept of two-body potential U(1,2).
Thus, in this context the answer is the familiar one.  The
force is the Coulomb force, known for 200 years, and for two point particles,
``1'' and ``2'', the associated potential is just the Coulomb potential
$U_C(1,2)$,
\begin{equation}
        U_C(1,2) = k_{12}/r, \hspace{1cm}  [k_{12} \equiv e_1e_2/4 \pi]
\end{equation}
with $r = |\bf r \rm _1-\bf r \rm _2|$; for two electrons, $k_{12}
\rightarrow  \alpha \equiv e^2/4\pi \approx
1/137$.
However, in quantum theory
the concept of force plays a secondary role and the question should really
be: ``What is the potential acting between two electrons?''  In the context of
nonrelativistic quantum mechanics (NRQM) and a configuration space description
of the system in question, the answer is again $U_C(1,2)$, now reinterpreted
as a multiplicative operator acting on a many-body Schrodinger wave function.
 
With regard to (ii), when the effects of the finite speed
of propagation of light are taken
into account, things get more complicated, both on conceptual and technical
grounds.  Now even in classical electrodynamics the question becomes murky:
While the meaning of the force on ``1'' is sharp, via the Lorentz force law,
the force exerted by ``2'' on ``1'' depends on the previous history of
``2'' (and
\it vice versa \rm ).  The question arises: To what extent can one describe the force
between ``1'' and ``2'' in terms of properties of their motion at a given instant?
This question appears to have been first addressed by C.E. Darwin [1], the
grandson of Charles Darwin. Darwin showed that, to order $v^2/c^2$, the effects
of retardation can be taken into account by adding to the free Lagrangian for
two point particles not only the Coulomb term -$U_C$ but also a term
-$U_D$, with
\begin{equation}
        U_D = -1/2(\bf v \rm _1 \cdot \bf v \rm _2+ \bf v \rm _1 \cdot
        \bf \hat{r}v \rm _2 \cdot \bf \hat{r} \rm )U_C/c^2,
\end{equation}
from which the corresponding force may be calculated.
On passing to the Hamiltonian formalism and introducing
canonical momenta one gets a momentum-dependent interaction term of the form 
\begin{equation}
        U_D = -1/2(\bf p \rm _1 \cdot \bf p \rm _2+ \bf p \rm _1 \cdot
        \bf \hat{r}p \rm _2 \cdot \bf \hat{r} \rm )U_C/m_1m_2c^2.
\end{equation}
This must have a counterpart in quantum theory.
But how is one to order the factors, if $\bf p \rm _i$ is replaced by
$\bf p \rm _i^{op}$?
In whatever way this question is answered, it becomes clear that at this
level any effective potential between electrons will not be strictly local.

Finally, with regard to (iii), after the discovery
of the spin-1/2 character of the electron and the associated magnetic
moment, it was obvious that spin-dependent terms must be added to $U_D$,
corresponding to the interaction of the magnetic field produced by ``1''
with the magnetic moment of ``2'' and \it vice versa \rm.  This led to the now familiar
spin-other-orbit and spin-spin potentials and gave rise to an operator
$U_{fs}^{(2)}$ which could describe the effect of electron-electron interaction
on atomic fine structure correctly to order $\alpha^2Ry$ (apart from a missing
contact term).  So one could say with some justification that, to order
$e^2$ and $v^2/c^2$, the operator $U_C+U_{fs}^{(2)}$ describes the force between
electrons.

After the development of the Dirac equation
one might have guessed that, within a framework
in which the state of a many-electron system is described by a multi-Dirac
spinor, the velocity factors $\bf v \rm_i$ in (2) should simply be replaced by
their formal counterparts in Dirac theory, \it viz \rm. $c \bf
\alpha \rm _i$.  This yields the operator
\begin{equation}
        U_B(1,2) = -1/2(\bf\alpha \rm_1 \bf \cdot \alpha \rm_2+ \bf
        \alpha \rm _1 \bf \cdot \hat {r} \alpha \rm _2 \bf \cdot \hat {r} \rm )U_C,
\end{equation}
known as the Breit operator.  It was first obtained
by Breit [2], but not in the way I have described.  After the development
of QED by Dirac, Breit studied the level shift in the helium atom arising
from the exchange of a transverse photon between the two electrons, within a
formally relativistic (but actually unsound) framework.  He concluded that, to
a good approximation, this shift is given by the expectation value of
$U_B$ with
a Dirac-type wave function which includes the effect of $U_C$.  From this one
would expect that an effective interaction between electrons which takes into
account items (i), (ii), and (iii) is given by the Coulomb-Breit potential,
\begin{equation}
        U_{CB}(1,2) = U_C(1,2) + U_B(1,2).
\end{equation}
However, although accepted for many years,
this expectation turns out to be misleading at best and wrong at worst.

In the following sections I will consider the
question of the electron-electron interaction from the
perspective of modern QED.  It will be useful to consider both spin-1/2 and
spin-0 particles. Sec. II deals with the relatively simple case of one-photon
exchange potentials, in the context of the scattering of two point particles.
However, 200 years after Coulomb one ought to have some idea about the extent
to which the exchange of two photons can also be represented by an effective
potential.  This is discussed  Sec. III.  In Sec. IV I turn to the question
of the connection between these potentials and the properties of bound states.

\section{Scattering potentials: One-photon-exchange}

Within the framework of perturbative
quantum field theory, there is a sharp contrast between the beautiful
methods available for the calculation of collision amplitudes and those
used in practice for the calculation of the properties of composite systems
or ``bound states,'' especially in the case of a gauge theory such as QED.
As an example, consider the scattering of two particles, ``1'' and ``2'',
\begin{equation}
        1 + 2 \rightarrow 1 + 2,
\end{equation}
with initial and final four-momenta $p_1$, $p_2$ and $p'_1 $, $p'_2$,
respectively. In a Lorentz-invariant
theory the transition amplitude has the form $t_{fi} = N_fMN_i$, where
$N_f = (4E_1'E_2')^{-1/2}$ and $N_i = (4E_1E_2)^{-1/2}$ are the usual kinematic
factors, and
\begin{equation}
        M = M(s,t) \hspace{1cm}    [s \equiv P^2, t = Q^2]
\end{equation}
is the invariant Feynman amplitude, with \it s \rm and \it t \rm the invariant squares
of the total four-momentum $P = p_1+p_2$ and four-momentum transfer $Q =
p_1-p_1'$.
Let us first consider the effects arising from one-photon exchange
between two distinct point-like spin-1/2 particles, such as an $e^-$ and a
$\mu^+$.

\subsection{Two spin-1/2 particles}

Using Feynman gauge for the photon propagator one finds that the
lowest-order contribution to $M$, regarded as a power series in $e_1e_2$, is
given by
\begin{equation}
        M^{(2)} = \bar{u}'_1  \bar{u}'_2 F^{(2)}u_1u_2
\end{equation}
with the $u_i$ denoting lepton spinors normalized to $2m_i$ and
\begin{equation}
        F^{(2)} = e_1e_2\gamma_1 \cdot \gamma_2/t.
\end{equation}
In the c.m. system, $Q \rightarrow (0, \bf q \rm )$, with \bf q \rm the three-momentum
transfer, $ \bf q = p-p' \rm $.
On taking the Fourier transform of $F_{cm}^{(2)}$ with a factor $ exp(-i
\bf q \cdot r \rm ) $,
one sees that the corresponding transition amplitude $t^{(2)}$ can be obtained
by taking the matrix element between plane wave spinors of the operator
\begin{equation}
        U_{CG}(1,2) = k_{12}(1-\bf \alpha \rm _1 \bf \cdot \alpha \rm _2)/r =  U_C(1,2)+U_G(1,2)
\end{equation}
with $U_G \equiv -k_{12}(\bf \alpha \rm _1 \bf \cdot \alpha \rm _2/r)$, the
so-called Gaunt potential [3].

It is therefore tempting to say that, to leading order in $e_1e_2$, but
regardless of the relative speed, the effective interaction between,
``1'' and
``2'' is given by (10).  However, suppose that instead one uses the Coulomb
gauge for the photon propagator.  Then one finds that (9) is replaced by
\begin{equation}
        F_T^{(2)} =
        e_1e_2[\gamma_1^0\gamma_2^0/Q^2+(\gamma_1 \cdot
        \gamma_2-\gamma_1 \cdot Q
        \gamma_2 \cdot Q/Q^2)/Q^2].
\end{equation}
Fourier transformation of $F_{T;cm}^{(2)}$ yields
\begin{equation}
        U_{CB}(1,2) = U_C+U_B,
\end{equation}
where $U_B$ is the operator
\begin{equation}
        U_B = -k_{12}( \bf \alpha \rm _1 \bf \cdot \alpha \rm _2+ \bf
        \alpha \rm _1 \bf \hat{r} \cdot \alpha \rm _2 \bf \cdot \hat{r}
        \rm )/2r
\end{equation}
originally obtained by Breit in the
context of a bound state problem, as mentioned above.
By construction, the Coulomb-Gaunt potential $U_{CG}$ and the Coulomb-Breit
potential $U_{CB}$ have the same matrix element between product plane wave
spinors,
provided these are on the energy shell $(\bf p \rm ^2 = \bf p \rm '^2)$,
a minimal requirement
for an effective potential.  Thus even in the simplest circumstances,
asking for ``the effective electron-electron potential'' is too naive.

More serious is the fact that neither $U_{CB}$ or $U_{CG}$ are
permissible approximations to an effective potential, in the context
of a Dirac-spinor description of spin-1/2 particle wave functions.  This
can be most easily seen by including a third particle ``3''.  The use of a sum
such as $U(1,2)+U(1,3)+U(2,3)$ to describe the interaction is disastrous from
the start, because of the problem of continuum dissolution [4,5].  Analysis
shows that a theoretically well-founded choice for the second-order potential
is neither $U_{CG}$ nor $U_{CB}$, but either of the two operators
$V_{CG}$ or
$V_{CB}$, defined by
\begin{equation}
        V_{CG} = \Lambda_{++}U_{CG}\Lambda_{++}, \quad V_{CB} =
        \Lambda_{++}U_{CB}\Lambda_{++},
\end{equation}
where $\Lambda_{++} = \Lambda_+(1)\Lambda_+(2)$ is the product of positive-energy
Casimir-type projection operators
for the leptons [5,6].  Since the spinors in (8) are eigenfunctions
of the $\Lambda_+(i)$, these operators also reproduce the lowest-order amplitude.

All this is by now fairly well known.  But a related question has received
relatively little attention:  What is an effective potential which describes
the scattering amplitude correctly to fourth order?  Before considering this,
it is useful to study two simpler but instructive cases: two spin-0 particles,
in scalar QED, and a spin-1/2 and a spin-0 particle, in spinor-scalar QED.

\subsection{Two spin-0 particles}

On using Feynman gauge to write down the one-photon exchange amplitude, one
gets
\begin{equation}
        M^{(2)} = e_1e_2(p_1+p_1') \cdot (p_2+p_2')/t =  e_1e_2(2a+t)/t,
\end{equation}
where $a = s-m_1^2-m_2^2$.  Fourier transformation of the c.m.value of (2.10)
yields a term proportional to $U_C$, with an energy-dependent coefficient,
plus a contact term proportional to $\delta(\bf r \rm)$.  Such a potential is
not suitable
for use in a Schr\"{o}dinger type of equation.  In second-order perturbation
theory it would lead to an ultraviolet (UV) divergence.  A potential
which is iterable can be obtained by first writing $M^{(2)}$ in a different
form (which does not change its value on the mass shell) and then finding
an equivalent operator in r-space, now involving derivative operators [7].
One is thereby led to what can be termed a Feynman-gauge-inspired
(FGI) potential $V_{1\gamma}^{FGI}$, which in the c.m. system $ ( \bf p
\rm _1^{op}\rightarrow  \bf p \rm _{op}, \bf p \rm _2^{op} \rightarrow - \bf p \rm _{op}) $ is
\begin{equation}
        V_{1\gamma} ^{FGI} =  z'_{op}U_Cz'_{op} +
        y_{op}( \bf p \rm _{op}ùU_C \bf p \rm _{op}/2m_Am_B)y_{op},
\end{equation}
with $z'_{op} \equiv (1+ \bf p \rm _{op}^2/2E_1^{op}E_2^{op})^{1/2}$ and $y_p =
(m_1m_2/E_1^{op}E_2^{op})^{1/2}$.
The corresponding Coulomb-gauge inspired (CGI) one-photon exchange
potential $V_{1\gamma}^{CGI}$ is given by [8],
\begin{equation}
        V_{1\gamma} ^{CGI} = V_C+V_T
\end{equation}
where, with a curly bracket denoting an anticommutator,
\begin{equation}
        V_C = y_{op}[\{E_1^{op},\{E_2^{op},U_C\}\}]y_{op}/4m_1m_2
\end{equation}
is a relativistic version of the Coulomb interaction and
\begin{equation}
        V_T = -
        (1/2)y_{op}\{ \bf p \rm _{1;i}^{op},\{ \bf p \rm
        _{2;j}^{op},(\delta_{ij}+\hat{r}_i\hat{r}_j)U_C\}\}]y_{op}/4m_1m_2
\end{equation}
is the potential arising from the exchange of a transverse photon.  

\subsection{One-photon exchange potential in spinor-scalar QED}

To complete the one-photon story, consider the exchange of a photon when
``1'' has spin-0 and ``2'' has spin-1/2.  The Feynman amplitude is then given by
\begin{equation}
        M^{(2)} = -e_1e_2\bar{u} \prime \gamma_2 \cdot (p_1+p_1 ')u/t
\end{equation}
The corresponding FGI potential is [9]
\begin{equation}
        V_{1\gamma}^{FGI} =
        y_1^{op}\Lambda_+^{op}(2)U^{(2)}\Lambda_+^{op}(2)y_1^{op} \quad
        [y_1^{op} \equiv (m_1/E_1^{op})^{1/2}]
\end{equation}
where
\begin{equation}
        U^{(2)} = (E_1^{op}U_C+U_C^{op}E_1^{op} - \bf \alpha \cdot p \rm
        _{op}U_C-U_C \bf \alpha \cdot p \rm _{op})/2m_1.
\end{equation}
This is an analogue of $V_{1\gamma}^{FGI}$ for two spin-1/2 particles.
There is a corresponding CGI
potential which I do not write down.  Use will be made of (21) in Sec. III.

\subsection{Remarks on the orbit-orbit interaction}
 
The difference between the two choices, $V_{1\gamma}^{FGI}$ and
$V_{1\gamma}^{CGI}$
is connected with the form of the so-called orbit-orbit
interaction.  To see this, note that in the n.r. limit (16) yields
as the leading correction to $U_C$ an orbit-orbit interaction $U_{o-o}$ of the form
\begin{equation}
        U_{o-o}^{FGI} = \{ \bf p \rm _i^{op},\{ \bf p \rm _j^{op},\delta_{ij}U_C\}\}/4m_Am_B,
\end{equation}
whereas (17) yields
\begin{equation}
        U_{o-o}^{CGI} = (1/2)\{\bf p \rm _i^{op},\{\bf p \rm
        _j^{op},(\delta_{ij}+\hat{r}_i\hat{r}_j)U_C\}\}/4m_Am_B.
\end{equation}
Note that (24) is a manifestly hermitian form of the orbit-orbit
interaction familiar from atomic physics, usually
described in texts as resulting from the reduction of the Breit operator
(13)
to n.r. form.  This is unfortunate from a pedagogical point of view,  since
the Breit operator refers only to spin-1/2 particles.  Spin has nothing to do
with it!  I will return to the difference between (23) and (24) shortly.

\section{Scattering potentials: Two-photon exchange}
	
A key point in the definition of any potential, often overlooked,
is that one must specify in advance how it
is to be used.  Let us work in the c.m. system of the particles, scattering
with total energy E.  With $h_0 = h_0(1)+h_0(2)$ the free Hamiltonian for
the particles, we will require that the transition amplitude computed from 
\begin{equation}
        T_{op}(E) =  V_{eff} + V_{eff}(E-h_0-V_{eff}+i\epsilon)^{-1}V_{eff}
\end{equation}
reproduce the scattering amplitude obtained from field theory to a given
accuracy.  I consider first the case of two spin-0 particles, then the mixed
case of a spin-0 and a spin-1 particle, and finally the case of two spin-1/2
particles.

\subsection{Two spin-0 particles}

In scalar QED there are five Feynman diagrams corresponding to
two-photon exchange, which contribute to the fourth-order
amplitude $M^{(4)}$: the two-rung ladder (or "box") graph, and
the two-rung crossed-ladder graph, a pair of "single-seagull" graphs, and a
"double-seagull" graph. The sea-gull graphs are ultraviolet (UV) divergent,
but these divergences are taken care of by renormalization and do not affect
the long-distance behavior.  More serious is the fact that the "box" and
"crossed box" graphs are infrared (IR) divergent.  Thus it seems at first sight
that one cannot even begin to talk about a two-photon exchange potential!

The resolution of the IR problem lies in the recognition that
the iteration amplitude $M_I$ arising from $V_{1\gamma}$ is now also
infrared divergent.
This divergence is just the field-theoretic counterpart of the fact
that in NRQM the second (and higher) Born approximations to the scattering
of a charged particle in a Coulomb field is divergent (for any value
of the momentum transfer).  It turns out that the difference
$M_{2 \gamma}-M_I$ is IR
convergent, so that the associated potential $V_{2\gamma}$ is
well-defined after all.

I will only state the form of the result at large \it r \rm (that
is, \it r \rm large
compared to the Compton wave length of either particle) and at low
momenta.  On using $V_{1\gamma}^{FGI}$ to compute $M_I$ one finds that [7],
\begin{equation}
        V_{2\gamma}^{FGI} = c_2^{FGI}r^{-2} + c_3^{FGI}r^{-3} + ...
\end{equation}
where, with $k \equiv e_1e_2/4 \pi$,
\begin{equation}
        c_2^{FGI} = k^2/2(m_1+m_2), \hspace{1cm} c_3^{FGI} = - 7k^2/6\pi
        m_1m_2.
\end{equation}
In contrast, use of $V_{1\gamma}^{CGI}$ yields [8]
\begin{equation}
        V_{2\gamma}^{CGI} = c_2^{CGI}r^{-2} + c_3^{CGI}r^{-3} + ...
\end{equation}
where
\begin{equation}
        c_2^{CGI} = 0, \hspace{1cm} c_3^{CGI} = - 7k^2/6 \pi m_1m_2.
\end{equation}

The difference between the asymptotic forms (25) and (28) can be traced
back the difference in the associated forms of the orbit-orbit interactions
mentioned above.  Thus we see that in the case of two charged particles the
leading asymptotic behavior of $V_{2\gamma}$ depends on the precise definition
of $V_{1\gamma}$.
This observation resolves a long-standing puzzle concerning conflicting
results for the value of $c_2$. Further, as was noted some time ago by L. Spruch,
$c_2^{FGI}$ is classical in character, i.e. if \it h \rm and
\it c \rm are restored,
$c_2$ turns out
to be independent of \it h \rm.  One should therefore try to understand the source of
this term from classical electrodynamics.  It turns out that this is indeed
possible by a reexamination of the work of Darwin [1], but I will not enter
into the details here [10].

\subsection{A spin-1/2 and a spin-0 particle}

A similar analysis can be carried out for the mixed case. If one uses
$V_{1\gamma}^{FGI}$ to compute the iteration amplitude one finds that the spin-
independent part of $V_{2\gamma}$ is
\begin{equation}
        V_{2\gamma}^{s.i.} = k^2[2(m_1+m_2)]^{-1}r^{-2},
\end{equation}
at large \it r \rm and low energies, which coincides with that for two spin-zero
particles.  The spin-dependent part is essentially a spin-other-orbit type of
interaction [9]
\begin{equation}
        V_{2\gamma}^{s.o.} =
        -(e_1e_2/4\pi)^2(\sigma \cdot \ell /4m_2^2)
        [(3m_1+5m_2)/m_1(m_1+m_2)]r^{-4}
\end{equation}
The two-photon-exchange spin-orbit interaction therefore decreases as
$r^{-4}$, or
one power more rapidly than that arising from one-photon exchange.
Applications of this result to a variety of exotic systems, such as pionium
and muonic helium, are considered in Ref. [9].

\subsection{Two spin-1/2 particles}

The computation of $V_{2\gamma}$ for two spin-1/2 particles, even of its
asymptotic form for large {\it r}, turns out to be remarkably complex and has not yet
been completed.  However, some aspects are known and others can be guessed.
For example, one can show that if $V_{1\gamma}^{FGI}$ is used for the lowest-order
potential, then the spin-independent part of $V_{2\gamma}$  coincides, for large
\it r \rm ,  with that for two spin-0 particles and the spin-other-orbit part coincides
with that for a spin-0 particle ``1'' and a spin-1/2 particle ``2'', plus a similar
term with ``1'' and ``2'' interchanged.  At large \it r \rm and low energies one expects
that the spin-spin part $V_{2\gamma}^{s-s}$ has the form
\begin{equation}
        V_{2\gamma}^{s-s} = k^2[A(r) \bf \sigma \rm_1 \bf \cdot \sigma
        \rm _2 + B(r) \bf \sigma \rm _1 \bf \cdot \hat{r} \sigma_1 \cdot
        \sigma_2 \cdot \hat{r} \rm ],
\end{equation}
with $A(r) = ar^{-p}$ and $B(r) = br^{-q}$.  The coefficients
\it a \rm and \it b \rm, and the
exponents
\it p \rm and \it q \rm , remain to be determined.  For equal-mass particles,
such as two
electrons, $a = a'/m^{p-1}$ where $a'$ is dimensionless and
likely to be of order
unity.  To guess the value of \it p \rm, note that in the large
\it r \rm, low-energy limit the
spin-spin part coming from $V_{1\gamma}$ has one more inverse power of
\it r \rm than the spin-
other-orbit part.  If the same relation holds for two-photon exchange, then
$p = 5$.

\section{Scattering potentials and bound states}

Let us now consider the possible relevance of all this to the computation of
relativistic atomic structure, one of Ingvar Lindgren's keen interests.

The application of QED to compute such structure to an accuracy beyond order
$\alpha^2Ry$, for systems more complicated than the classic two-body systems (hydrogen,
positronium, and muonium) began in the mid-fifties, with calculations of the
fine structure of helium to an accuracy of order $\alpha^3Ry$ [11,12].  These
calculations, and subsequent ones of spin-dependent effects of order
$\alpha^4Ry$ [13],
were based on a generalization of the two-body Bethe-Salpeter (BS) equation to include
an external field. The generalized BS equation leads in a somewhat convoluted
way to an equation of the form
\begin{equation}
        H_{++}\Psi = E\Psi
\end{equation}
where
\begin{equation}
        H_{++} = h_D^{ext}(1) + h_D^{ext}(2) + L_{++}U_C(1,2)L_{++},
\end{equation}
with   $h_D^{ext}(i) = \bf \alpha \rm _i \cdot \bf p \rm _i+\beta_im-Z\alpha/r_i$ the usual external-field Dirac Hamiltonian
and $L_{++}$ the product of the associated external-field positive-energy projection
operators.  This equation, called the (external-field) no-pair Coulomb-ladder
equation in [12], can be used as the starting point of a systematic perturbative
approach to the energy levels of He and He-like systems.  The corresponding
equation with $U_C$ replaced by $U_C+U_B$ contains not only all $\alpha^2Ry$ corrections to
fine-structure, but a number of corrections of order $\alpha^3Ry$ as well.  (For large Z
these can be expressed in analytic form. Recently, Lindgren and his collaborators
[14] have developed methods for solving such equations numerically with
sufficient accuracy to enable them to verify the analytic results.)

However, the further calculations needed to obtain the remaining
$\alpha^3Ry$ and still
higher-order corrections, such as those coming from two-photon exchange, are
complicated, messy, and have all the appeal of a thorny, black box.  They
contrast strongly with the beauty of scattering-amplitude calculations, where
among other things one can advantageously use a covariant propagator for the
photon.  It is an attractive idea to try to understand what part of the level
structure can be thought of as arising purely from those forces which act when
the electrons undergo free scattering.  To put it another way, note that in
NRQM one can describe both the scattering of two electrons and their interaction
within a bound state with one and the same interaction potential $U_C$.  The
question then is to what extent one can, within the framework of QED, understand
the bound states in terms of the same two-body effective potentials which
describe the scattering accurately.  It is clear that in QED there are effects
which cannot be so described, \it e.g. \rm the effects of the external potential
provided by the nucleus during the exchange of photons between the electrons.

The availability of a lepton-lepton potential which reproduces the scattering
amplitude exactly to order $(e_1e_2)^2$ would be a significant step in the
direction of such a goal [15].  An early test of such a program would be
provided, for example, by solution the ``free no-pair ladder equation'' [5]
\begin{equation}
        h_{++}\Psi = E\Psi
\end{equation}
where
\begin{equation}
        h_{++} = h_+^{ext}(1) + h_+^{ext}(2) + V_{++}.
\end{equation}
Here $h_+^{ext}(i) = \bf \alpha \rm _i \bf \cdot p \rm _i+\beta_{i}m +
\Lambda_+(i)(-Z\alpha/r_i)\Lambda_+(i)$, and $V_{++}$ is of the form
$V_{++} =  \Lambda_{++}U(1,2)\Lambda_{++}$, with $\Lambda_{++}$ the product
of free positive-energy projection
operators, as in (14).  The operator $U(1,2) = U^{(2)} + U^{(4)}$
must be chosen so
that if the external potential is turned off, the one-photon and two-photon
exchange scattering amplitudes are reproduced. (I note in passing that if one
uses $U^{(2)} = U_{CG}$ rather that $U_{CB}$ one must already include
$U^{(4)}$ to get the correct
$\alpha^2Ry$ level shift!)  On comparing the eigenvalues of (35) with those obtained
from, e.g., the Bethe-Salpeter equation approach, one could disentangle those
dynamical effects involved in the scattering from those which are peculiar to
the bound-state situation.  For \it Z \rm not too large one would have an initial
description in the language of configuration space which is not only quite
accurate, but which also could actually be explained to students, without
requiring that they become experts in the arcana of four-dimensional wave
equations and all that.  So it may well appeal to Ingvar Lindgren:  As the
photo-op at the end of this symposium showed, the results of his pedagogical
skill and devotion could only be captured by a wide-angle lens!

\begin{center}
{\bf Acknowledgements}
\end{center}

This work was supported in part by the U.S. National Science Foundation.
\begin{center}

\newpage
{\bf References}
\end{center}
\begin{enumerate}
\item \label{list:one} C.E. Darwin, Philos. Mag. \bf 39 \rm, 537 (1920).
\item \label{list:two} G. Breit, Phys. Rev. \bf 34 \rm , 553 (1929).
\item \label{list:three} J.B. Gaunt, Proc. R. Soc. London, Ser. A \bf 122 \rm, 513 (1929).
\item \label{list: four} G. Brown and D.G Ravenhall, Proc. R. Soc. London, Ser. A \bf 208 \rm, 552 (1951).
\item \label{list:five} J. Sucher, Phys. Rev. A \bf 25 \rm , 348 (1980); Phys. Rev. Lett. \bf 55 \rm, 1023 (1985).
\item \label{list:six} For a review, see e.g. J. Sucher, in Program on Relativistic, Quantum
   Electrodynamic and Weak Interaction Effects in Atoms, edited by W.R.
   Johnson, P. Mohr, and J. Sucher (AIP, New York, 1989).
\item \label{list:seven} G. Feinberg and J. Sucher, Phys. Rev. D \bf 38 \rm, 3763 (1988).
\item \label{list:eight} J. Sucher, Phys. Rev. D \bf 49 \rm , 4284 (1994).
\item \label{list:nine} G. Feinberg and J. Sucher. Phys. Rev. D  \bf 45 \rm, 2493 (1992).
\item \label{list:ten} J. Sucher, Comm. At. Mol. Phys. \bf 30 \rm , 129 (1994).
\item \label{list:eleven} H. Araki, Prog. Theoret. Phys. \bf 17 \rm , 6191 (1957).
\item \label{list:twelve} J. Sucher, Phys. Rev. \bf 109 \rm , 1010 (1958).
\item \label{list:thirteen} M.H. Douglas and N.M. Kroll, Ann. Phys. (N.Y.) \bf 82 \rm , 89 (1974)
\item \label{list:fourteen} I. Lindgren et al., Phys. Rev. \bf A51 \rm , 1167 (1995).
\item \label{list:fifteen} There is an additional force between leptons, arising from the exchange of
    a neutrino-antineutrino pair, which is unfortunately too weak to be
    accessible experimentally.  For a review, see J. Sucher, in \it Proceedings of
    the XIVth Moriond Workshop: Particle Astrophysics, Atomic Physics and
    Gravitation \rm , Villars-sur-Ollon, January, 1994.

\end{enumerate}
\end{document}